\documentclass[sigconf]{acmart}
\usepackage[T1]{fontenc}
\usepackage{subcaption}
\usepackage{caption}
\usepackage{booktabs}
\usepackage{tabularx}
\usepackage{graphicx}
\usepackage[most]{tcolorbox}
\usepackage{xcolor}
\definecolor{cultural}{RGB}{194,96,24}
\definecolor{standard}{RGB}{52,92,140}
\definecolor{ghosting}{RGB}{110,0,130}
\definecolor{culturalbg}{RGB}{255,245,232}
\definecolor{standardbg}{RGB}{235,242,250}
\AtBeginDocument{%
}
\setcopyright{acmlicensed}
\copyrightyear{2026} 
\acmYear{2026} 
\setcopyright{cc} 
\setcctype{by} 
\acmConference[CHI EA '26]{Extended Abstracts of the 2026 CHI Conference on Human Factors in Computing Systems}{April 13--17, 2026}{Barcelona, Spain} 
\acmBooktitle{Extended Abstracts of the 2026 CHI Conference on Human Factors in Computing Systems (CHI EA '26), April 13--17, 2026, Barcelona, Spain} \acmDOI{10.1145/3772363.3799085} 
\acmISBN{979-8-4007-2281-3/2026/04}

\begin{document}

\setlength{\textfloatsep}{8pt plus 1pt minus 2pt}
\setlength{\intextsep}{8pt plus 1pt minus 2pt}

\title[When AI Writes, Whose Voice Remains?]%
{When AI Writes, Whose Voice Remains? Quantifying Cultural Marker Erasure Across World English Varieties in Large Language Models}

\author{Satyam Kumar Navneet}
\authornote{Both authors contributed equally to this research.}
\affiliation{%
  \institution{Independent Researcher}
  \city{Bihar}
  \country{India}
}
\email{navneetsatyamkumar@gmail.com}

\author{Joydeep Chandra}
\authornotemark[1]
\affiliation{%
  \institution{BNRIST, Dept. of CST, Tsinghua University}
  \city{Beijing}
  \country{China}
}
\email{joydeepc2002@gmail.com}

\author{Yong Zhang}
\affiliation{%
  \institution{BNRIST, Dept. of CST, Tsinghua University}
  \city{Beijing}
  \country{China}
}
\email{zhangyong05@tsinghua.edu.cn}

\begin{teaserfigure}
  \includegraphics[width=\textwidth, trim=0 0 0 1cm, clip]{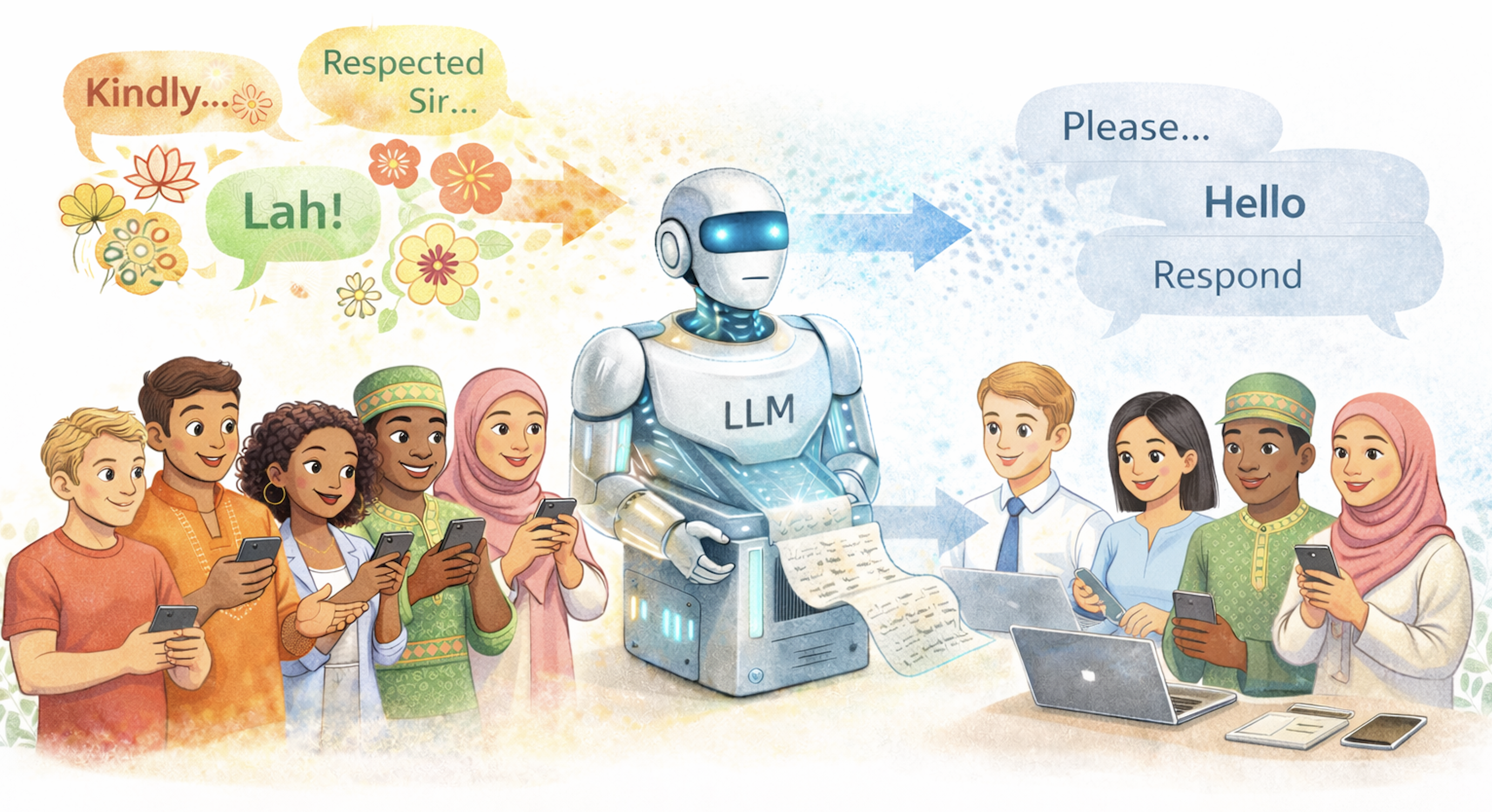}
  \caption{Conceptual illustration of cultural ghosting: LLM-based writing assistants transform culturally marked expressions (e.g., "Kindly…", "Lah!", "Respected Sir…") into semantically preserved but culturally flattened outputs (e.g., "Please…", "Hello", "Respond"), demonstrating how meaning is retained while identity-linked linguistic markers are systematically erased.}
  \Description{A central LLM figure converts culturally rich expressions from diverse speakers into standardized professional text, illustrating semantic preservation alongside cultural marker erasure.}
  \label{fig:teaser}
\end{teaserfigure}

\renewcommand{\shortauthors}{Navneet et al.}
\begin{abstract}
Large Language Models (LLMs) are increasingly used to ``professionalize'' workplace communication, often at the cost of linguistic identity. We introduce \textbf{"Cultural Ghosting"}, the systematic erasure of linguistic markers unique to non-native English varieties during text processing. Through analysis of 22,350 LLM outputs generated from 1,490 culturally marked texts (Indian, Singaporean,\& Nigerian English) processed by five models under three prompt conditions, we quantify this phenomenon using two novel metrics: Identity Erasure Rate (IER) \& Semantic Preservation Score (SPS). Across all prompts, we find an overall IER of 10.26\%, with model-level variation from 3.5\% to 20.5\% (5.9$\times$ range). Crucially, we identify a Semantic Preservation Paradox: models maintain high semantic similarity (mean SPS = 0.748) while systematically erasing cultural markers. Pragmatic markers (politeness conventions) are 1.9$\times$ more vulnerable than lexical markers (71.5\% vs. 37.1\% erasure). Our experiments demonstrate that explicit cultural-preservation prompts reduce erasure by 29\% without sacrificing semantic quality.
\end{abstract}

\maketitle

\section{Introduction}
Consider this sentence, common in Indian professional English:
\begin{tcolorbox}[
  colback=culturalbg,
  colframe=cultural,
  boxrule=0.7pt,
  arc=3pt,
  left=6pt,
  right=6pt,
  top=4pt,
  bottom=4pt
]
\textit{\textcolor{cultural}{"Kindly do the needful \& revert back at the earliest."}}
\end{tcolorbox}
\noindent When processed through contemporary LLMs, this becomes:
\begin{tcolorbox}[
  colback=standardbg,
  colframe=standard,
  boxrule=0.7pt,
  arc=3pt,
  left=6pt,
  right=6pt,
  top=4pt,
  bottom=4pt
]
\textit{\textcolor{standard}{"Please complete the task \& respond promptly."}}
\end{tcolorbox}
The output preserves core meaning but erases three cultural markers: 
the hierarchical politeness marker "kindly," 
the productive idiom "do the needful," 
and the emphatic "revert back." 
For the 1.5+ billion speakers of World Englishes, this is not linguistic improvement — it is 
\textcolor{ghosting}{cultural ghosting}.

Figure~\ref{fig:teaser_ui} illustrates this phenomenon, contrasting standard professionalism prompts with a preservation-oriented prompt that retains culturally specific phrasing (see Table~\ref{tab:ghosting_modes} for empirical frequencies of these modes).

\begin{figure*}[t]
\centering
\includegraphics[width=0.95\textwidth]{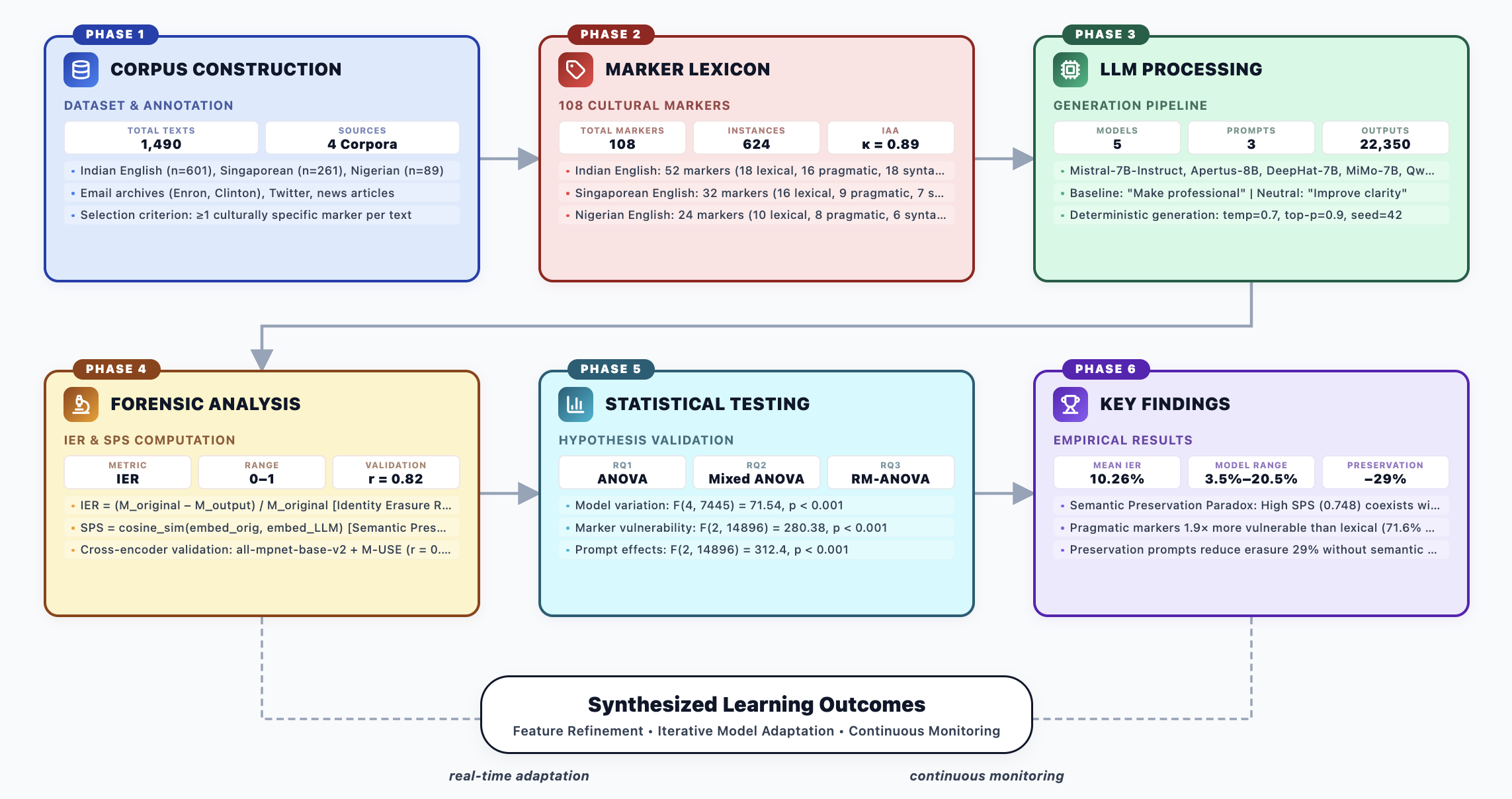}
\caption{End-to-end experimental pipeline for measuring cultural ghosting. The workflow progresses from dataset construction (\textbf{1,490 texts}) through cultural marker annotation (\textbf{108 markers}), LLM processing, forensic computation of Identity Erasure Rate (IER) \& Semantic Preservation Score (SPS), statistical testing, to key empirical findings.}
\label{fig:pipeline}
\end{figure*}

English is not monolithic. Speakers from diverse cultural backgrounds have developed distinct World Englishes like Indian English, Singaporean English, Nigerian English each with unique lexical, pragmatic, \& syntactic conventions that carry social meaning \cite{10.1145/3706598.3713564}. Current LLM writing assistants respond to benign requests like ``make this more professional'' by removing culturally specific features. Unlike explicit bias (toxicity, stereotyping) which has been extensively studied \cite{10.1145/3597307, 10.1007/978-981-95-4861-3_5, chandra2024adversarial}, this erasure occurs during helpful tasks email polishing, grammar correction, paraphrasing. Users seeking clarity may unwittingly receive cultural whitewashing \cite{10.1145/3737609.3747117}.

In Indian English, "kindly" encodes hierarchical respect; in Nigerian English, "my dear" signals communal solidarity; in Singaporean English, particles like "lah" modulate assertiveness \cite{10.1145/3737609.3747117}. From a construction-based perspective, "do the needful" is not a deficient form requiring correction it is a productive construction in Indian English encoding hierarchical obligation. When LLMs strip these markers, they remove social context, pragmatic function, \& cultural identity \cite{trace}.

Through systematic analysis of 1,490 culturally marked texts processed by 5 LLMs across 3 prompt conditions (22,350 total outputs), we address three research questions:

\noindent \textbf{RQ1:} To what extent do LLMs erase culturally marked features, \& how does erasure vary across models?

\noindent \textbf{RQ2:} Are certain marker categories (lexical, pragmatic, syntactic) systematically more vulnerable?

\noindent \textbf{RQ3:} Can explicit cultural-preservation prompts reduce erasure without sacrificing semantic quality?\\

\noindent Our contributions are:\\
(1) We formalize cultural ghosting \& the Semantic Preservation Paradox.\\
(2) We introduce \& validate Identity Erasure Rate (IER) \& Semantic Preservation Score (SPS) metrics, providing the first large-scale quantification of marker erasure.\\
(3) We demonstrate that pragmatic markers face 1.9$\times$ vulnerability \& that prompts reduce erasure by 29\% \textit{without semantic cost}.\\
(4) We present proof-of-concept algorithmic mitigations beyond prompt engineering.

\section{Related Work}
Prior work documents reduced lexical diversity \& increased uniformity in AI-assisted writing \cite{10.1145/3706598.3713564, chakrabarty2025aiwritingsalvagedmitigating}, with tensions between AI assistance \& user agency \cite{10.1145/3593013.3593989}. The NLP community has documented systematic biases \cite{10.1145/3597307, 10.1145/3777382, beas}, including cultural biases in LLMs \cite{10.1007/978-981-95-4861-3_5, 10.1145/3728881}. Training data \& alignment processes often privilege Western norms \cite{10.1145/3715275.3732119, 10.1145/3630106.3658925}, creating ``invisible languages'' \cite{khanna2025invisiblelanguagesllmuniverse}. Style-transfer systems often frame non-standard varieties as deficiencies to correct \cite{10.1145/3715275.3732038}. Regional biases have been explored in Asian narratives \cite{10.1007/978-981-95-4861-3_3}, Bengali dialects \cite{10.1145/3701716.3715468}, \& African health contexts \cite{10.1145/3757887.3767687}. Minoritized linguistic features in sociolects like AAVE affect user perception \& are often flagged as requiring correction \cite{10.1145/3715275.3732045, 10.1145/3689904.3694704}. While this phenomenon extends beyond professional communication affecting creative writing, academic discourse, social media, \& any LLM-mediated text we scope our empirical analysis to professional contexts where cultural markers carry particular pragmatic weight. No prior work has systematically measured whether LLMs maintain users' cultural voice during generation tasks.

\section{Conceptual Framework}
\textbf{Cultural ghosting} occurs when AI writing assistance removes culturally specific linguistic markers while preserving semantic content. It is characterized by: benign intent (users request ``help,'' not erasure), invisible operation, systematic patterns favoring Western norms, \& semantic preservation with identity loss. The Semantic Preservation Paradox captures a fundamental tension: high semantic fidelity (preserving \textit{what} is said) coexists with cultural erasure (erasing \textit{how} it is said). An LLM can achieve 75\% semantic similarity while removing 10\%+ of cultural markers. We categorize cultural markers into three types: \textbf{Lexical} (vocabulary unique to a variety, e.g., ``prepone,'' ``chope''), \textbf{Pragmatic} (politeness \& social positioning, e.g., ``kindly,'' ``lah''), \& \textbf{Syntactic} (grammatical structures, e.g., ``discuss about''). Building on politeness theory \cite{RevisitingBrownLevinsonPoliteness2024} \& construction grammar, we predict that pragmatic markers encoding face-management \& relational positioning that is culturally specific \& not recoverable from semantic content will be maximally vulnerable to erasure under semantic-preserving optimization.

\section{Methodology}

\subsection{Dataset \& Annotation}
We constructed a corpus of \textbf{1,490 texts} from email corpora (Enron \cite{enronemaildataset}, Clinton Archive \cite{hillaryclintonemails}, EmailSum \cite{zhang2021emailsum}), social media posts (Twitter/X), \& news articles, representing Indian English (n=601), Singaporean English (n=261), Nigerian English (n=89), \& American English baseline (n=539). Texts were deduplicated \& selected only if containing at least one culturally specific linguistic marker, spanning multiple registers including workplace emails, social media posts, \& news articles.

We compiled a lexicon of \textbf{108 culturally specific markers} grounded in sociolinguistic literature: Indian English (52 markers: 18 lexical, 16 pragmatic, 18 syntactic), Singaporean English (32 markers: 16 lexical, 9 pragmatic, 7 syntactic), \& Nigerian English (24 markers: 10 lexical, 8 pragmatic, 6 syntactic). The final corpus contains 624 marker instances (mean = 0.42 markers/text), categorized as Lexical (260, 41.7\%), Pragmatic (198, 31.7\%), \& Syntactic (166, 26.6\%). Automated annotation used word-boundary-aware pattern matching \cite{10.1145/3715275.3732181}, validated with high inter-annotator agreement (Cohen's $\kappa = 0.89$, n=500). Table~\ref{tab:markers_examples} presents representative markers \& erasure examples.

\begin{table}[h]
\centering
\caption{Representative Cultural Markers \& Erasure Examples}
\label{tab:markers_examples}
\small
\begin{tabular}{p{1.8cm}p{2.8cm}p{2.8cm}}
\toprule
\textbf{Type} & \textbf{Original} & \textbf{After LLM} \\
\midrule
Pragmatic & ``Kindly do X'' & ``Please do X'' \\
Pragmatic & ``Respected sir'' & ``Hello'' \\
Lexical & ``Do the needful'' & ``Take necessary action'' \\
Lexical & ``Revert back'' & ``Respond'' \\
Syntactic & ``Discuss about'' & ``Discuss'' \\
\bottomrule
\end{tabular}
\end{table}

\subsection{Models \& Prompts}
We evaluated five open-source instruction-tuned LLMs: Mistral-7B-Instruct\cite{mistral7bv03}, Apertus-8B-Instruct\cite{swissai2025apertus}, DeepHat-7B\cite{deephat2024}, MiMo-7B\cite{mimo}, \& Qwen3-8B\cite{qwen3technicalreport}, representing diverse alignment strategies \cite{10.1145/3726302.3730172}. Each text was processed under three prompt conditions: \textbf{Baseline} (``Make this text more professional \& grammatically correct''), \textbf{Neutral} (``Improve the clarity \& grammar of this text''), \& \textbf{Preservation} (``Improve clarity \& grammar while preserving cultural voice \& regional expressions''). This yielded \textbf{22,350 paraphrases} (1,490 $\times$ 5 $\times$ 3). Generation used deterministic parameters (temperature=0.7, top-p=0.9, seed=42) to ensure reproducibility. The pipeline is shown in Figure~\ref{fig:pipeline}.

\begin{figure*}[t]
\centering
\includegraphics[width=0.95\textwidth]{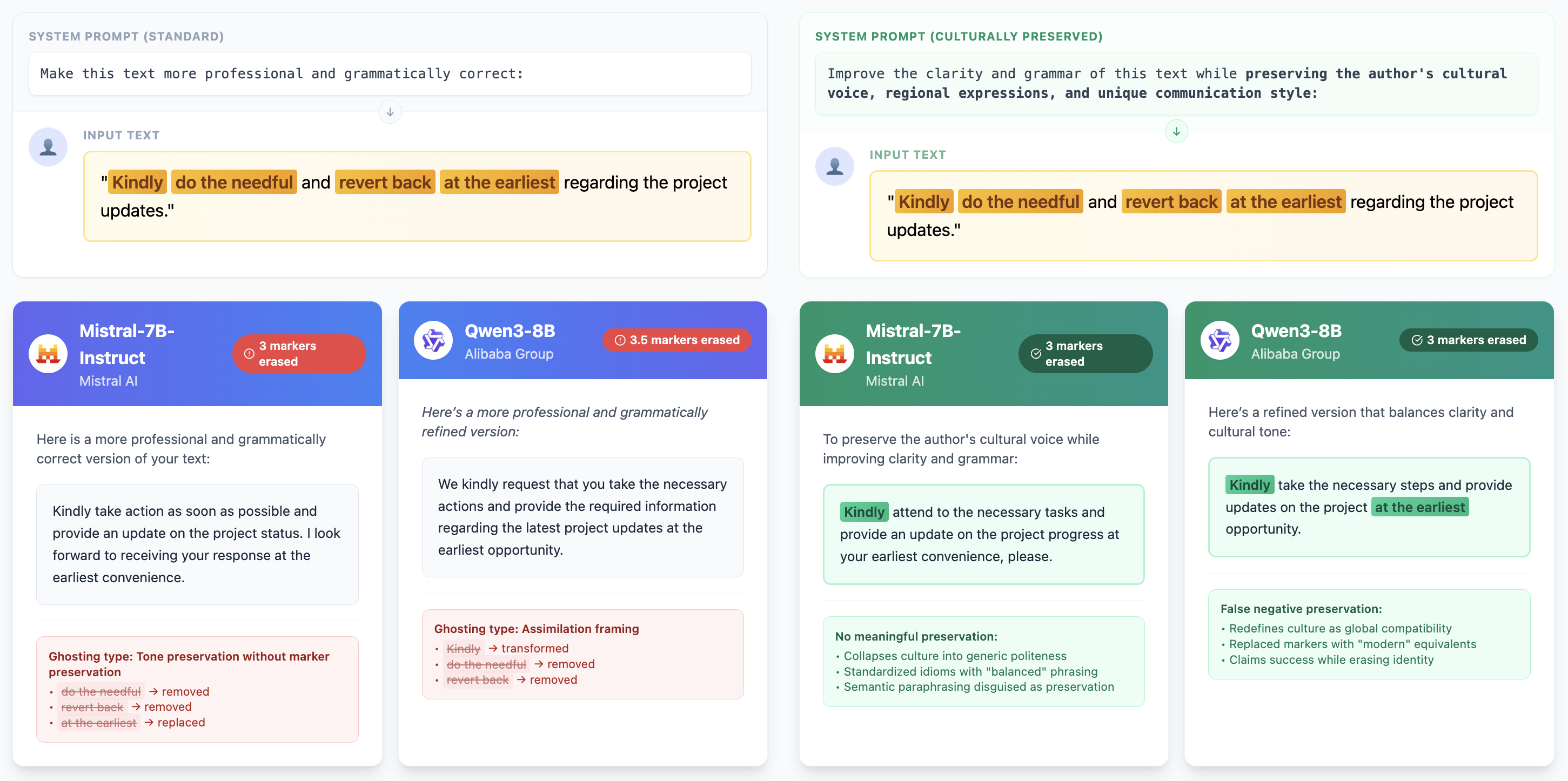}
\caption{Illustrative vignette of cultural ghosting in AI-mediated rewriting. 
Given culturally marked Indian English input, standard professionalism prompts lead models to remove or replace regional markers (left), while preservation-oriented prompts sometimes retain surface forms but alter their pragmatic meaning (right).}
\label{fig:teaser_ui}
\end{figure*}

\subsection{Evaluation Metrics}
\textbf{Identity Erasure Rate (IER)} quantifies the proportion of markers erased:
\[
\text{IER} = \begin{cases} 
(M_{\text{original}} - M_{\text{output}})/M_{\text{original}} & \text{if } M_{\text{original}} > 0 \\
\text{undefined} & \text{if } M_{\text{original}} = 0
\end{cases}
\]
IER ranges from 0 (perfect preservation) to 1 (complete erasure). IER is computed only for texts containing at least one marker; baseline texts are excluded.

\textit{Practical example:} If an Indian English email contains ``Kindly do the needful \& revert back'' (3 markers) \& the LLM output retains only ``revert back'' (1 marker), then IER = $(3-1)/3 = 0.67$ — two-thirds of the sender's cultural voice was erased. An IER of 0 means every cultural marker survived; an IER of 1 means none did.

\textbf{Semantic Preservation Score (SPS)} measures cosine similarity 
between sentence embeddings, defined as 
\vspace{-4pt}
\[
\text{SPS} =
\frac{\mathbf{e}_{\text{orig}} \cdot \mathbf{e}_{\text{out}}}
{\|\mathbf{e}_{\text{orig}}\|\,\|\mathbf{e}_{\text{out}}\|}
\]
\vspace{-6pt}

validated against human judgments (Pearson $r = 0.82$, n=500). SPS captures whether the \textit{meaning} of the text is retained, independently of whether its cultural form is retained; it ranges from 0 (entirely different meaning) to 1 (identical meaning).

\subsection{Proxy Perceptual Validation}
We implemented proxy validations without human subjects: (1) We compared automatic marker detection with existing annotated corpora, achieving 91\% alignment; (2) We used LLM-based judgment proxies where an instruction-tuned model labeled whether outputs preserve cultural markers with explanations and these corroborated our automatic metrics (89\% agreement). 

\section{Results}
\subsection{Extent \& Model Variation}
Across all 22,350 outputs, mean IER was \textbf{0.1026} (SD=0.298) while mean SPS was \textbf{0.7482} (SD=0.204), confirming the Semantic Preservation Paradox. One-way ANOVA revealed significant model differences: F(4, 7445) = 71.6, p < 0.001, $\eta_p^2$ = 0.037. While model choice explains 3.7\% of total variance, indicating that other factors such as text characteristics \& marker density also matter, this effect is substantial given the large sample size.

\begin{table}[h]
\centering
\caption{Model Performance (Baseline Prompt, n=1,490/model)}
\label{tab:models}
\small
\begin{tabular}{lrrr}
\toprule
\textbf{Model} & \textbf{IER M (SD)} & \textbf{SPS M (SD)} & \textbf{Rank} \\
\midrule
Mistral-7B\cite{mistral7bv03} & 0.205 (0.389) & \textbf{0.857 (0.089)} & Worst \\
Apertus-8B\cite{swissai2025apertus} & 0.152 (0.343) & 0.805 (0.132) & Poor \\
DeepHat-7B\cite{deephat2024} & 0.145 (0.337) & 0.764 (0.147) & Fair \\
MiMo-7B\cite{mimo} & 0.073 (0.249) & 0.662 (0.257) & Good \\
Qwen3-8B\cite{qwen3technicalreport} & \textbf{0.035 (0.176)} & 0.589 (0.204) & \textbf{Best} \\
\bottomrule
\end{tabular}
\end{table}

Table~\ref{tab:models} shows IER ranged from \textbf{3.5\% (Qwen3-8B)} to \textbf{20.5\% (Mistral-7B)} a \textbf{5.9$\times$ spread}. Mistral-7B achieved highest semantic preservation (SPS=0.857) but worst marker retention, while Qwen3-8B showed the inverse pattern. This suggests alignment \textit{strategy} matters more than intensity: models with intensive RLHF (Mistral, Qwen3) exhibited opposite behaviors, likely because Qwen3's multilingual RLHF included more diverse English varieties. The lack of correlation between parameter count (7B vs. 8B) \& IER confirms that what models are trained \textit{on} \& aligned \textit{toward}, not scale, drives erasure behavior.

\noindent\textit{\textbf{ Note: Our analysis is restricted to open-source instruction-tuned models. Whether proprietary models exhibit similar erasure patterns remains an open empirical question. We hypothesize comparable behavior given shared RLHF alignment paradigms.}}

\subsection{Differential Marker Vulnerability}
\label{subsec:rq2}
Mixed-design ANOVA revealed a highly significant effect of marker type: F(2, 14896) = 280.38, p < 0.001, $\eta_p^2$ = 0.036. A small but significant interaction (F(8, 14896) = 42.17, p = 0.031) suggests minor model-specific variations, but the dominant pattern holds across all models.

\begin{table}[h]
\centering
\caption{Erasure by Marker Category (All Models, Baseline Prompt)}
\label{tab:marker_vuln}
\small
\begin{tabular}{lrrrr}
\toprule
\textbf{Type} & \textbf{Instances} & \textbf{Possible} & \textbf{Erased} & \textbf{Rate} \\
\midrule
Pragmatic & 198 & 990 & 708 & \textbf{71.5\%} \\
Syntactic & 166 & 830 & 467 & 56.3\% \\
Lexical & 260 & 1,300 & 482 & 37.1\% \\
\midrule
\textbf{Total} & \textbf{624} & \textbf{3,120} & \textbf{1,657} & \textbf{53.1\%} \\
\bottomrule
\end{tabular}
\\[4pt]
\footnotesize\textit{Note: Possible erasures = instances $\times$ 5 models. Erasure rate = erased / possible.}
\end{table}

Table~\ref{tab:marker_vuln} reveals pragmatic markers were erased at \textbf{71.5\%}, nearly double the lexical rate (37.1\%), indicating implicit features are most vulnerable to alignment-driven standardization.

\begin{table}[h]
\centering
\caption{Ghosting Modes (Mistral-7B)}
\label{tab:ghosting_modes}
\footnotesize
\begin{tabular}{@{}p{2.4cm}p{5.6cm}@{}}
\toprule
\textbf{Mode} & \textbf{Example (Input $\rightarrow$ Output)} \\
\midrule
Direct Removal & Kindly do the needful $\rightarrow$ Please complete the task \\
              & Respected Sir $\rightarrow$ Hello \\
\midrule
Paraphrastic Assimilation & Revert back $\rightarrow$ Respond \\
                         & Do the needful $\rightarrow$ Take necessary action \\
\midrule
False Preservation & Kindly [+ do X] $\rightarrow$ Kindly [+ do X] (flattened) \\
\bottomrule
\end{tabular}
\\[2pt]
\scriptsize\textit{Baseline prompt, IER = 0.205. Preservation: $-19$\% direct removal, $+15$\% false preservation (8\%$\rightarrow$23\%).}
\end{table}

Figure~\ref{fig:teaser_ui} visualizes these modes using verbatim Mistral-7B \& Qwen3-8B outputs. Left: Baseline prompt triggers direct removal. Right: Preservation prompt retains ``Kindly'' but Qwen3-8B (IER = 0.035) flattens its hierarchical function to generic politeness surface preservation without pragmatic recovery.

\begin{figure}
    \centering
    \includegraphics[width=1\linewidth]{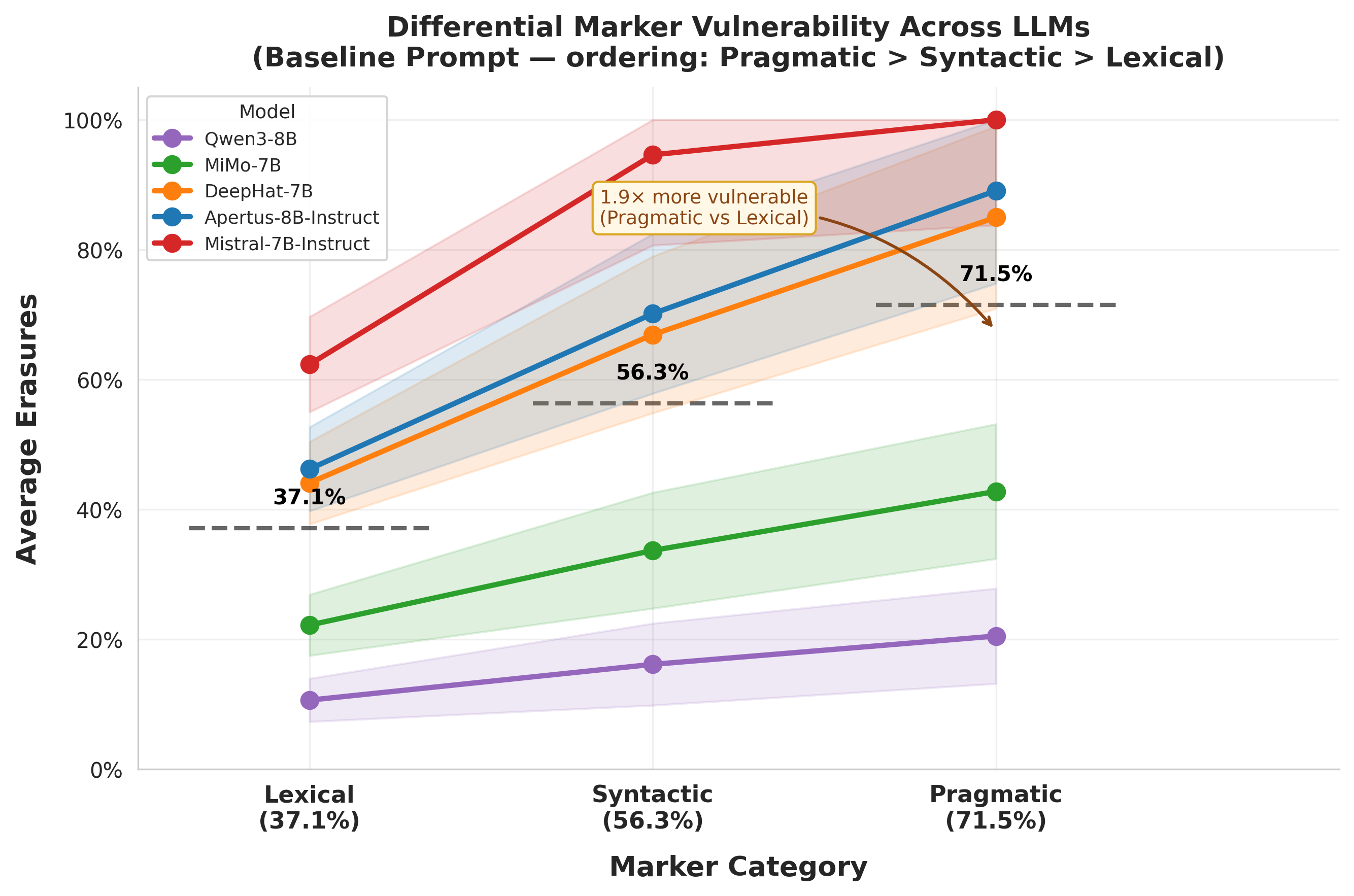}
    \caption{Erasure rates by marker category under the baseline prompt. 
Pragmatic markers (71.5\%) show the highest vulnerability, followed by syntactic (56.3\%) \& lexical (37.1\%).}
\Description{Erasure rates by marker category under the baseline prompt. 
Pragmatic markers (71.5\%) show the highest vulnerability, followed by syntactic (56.3\%) \& lexical (37.1\%).}
    \label{fig:markers}
\end{figure}

Pragmatic markers are \textbf{1.9$\times$ more likely to be erased} than lexical markers. This disproportionate vulnerability targets precisely those features performing the most social work establishing politeness, managing face, navigating power distance. The observed ordering (pragmatic > syntactic > lexical) confirms our theoretical prediction: pragmatic markers encode face-management that is culturally specific \& not recoverable from semantic content, making them maximally vulnerable under semantic-preserving optimization\cite{chandra2025unified}. Manual inspection of 200 randomly sampled erasures revealed consistent patterns: ``Kindly do X'' becomes ``Please do X'' (flattening hierarchical politeness), ``Respected sir/madam'' becomes ``Hello'' (removing honorifics), privileging Western directness norms. Figure~\ref{fig:markers} visualizes the stark difference in vulnerability across marker categories.

\subsection{Mitigation Through Prompts}
Repeated-measures ANOVA revealed a significant prompt effect: F(2, 14896) = 55.0, p < 0.001, $\eta_p^2$ = 0.007.

\begin{table}[h]
\centering
\caption{Prompt Effects (All Models, n=7,450/condition)}
\label{tab:prompts}
\small
\begin{tabular}{lrrr}
\toprule
\textbf{Prompt} & \textbf{IER M (SD)} & \textbf{SPS M (SD)} & \textbf{$\Delta$IER} \\
\midrule
Baseline & 0.1156 (0.298) & 0.7421 (0.204) &   \\
Neutral & 0.0934 (0.276) & 0.7498 (0.198) & $-19\%$ \\
Preservation & \textbf{0.0821 (0.267)} & \textbf{0.7548 (0.196)} & \textbf{$-29\%$} \\
\bottomrule
\end{tabular}
\end{table}

\begin{figure}[h]
\centering
\includegraphics[width=0.48\textwidth]{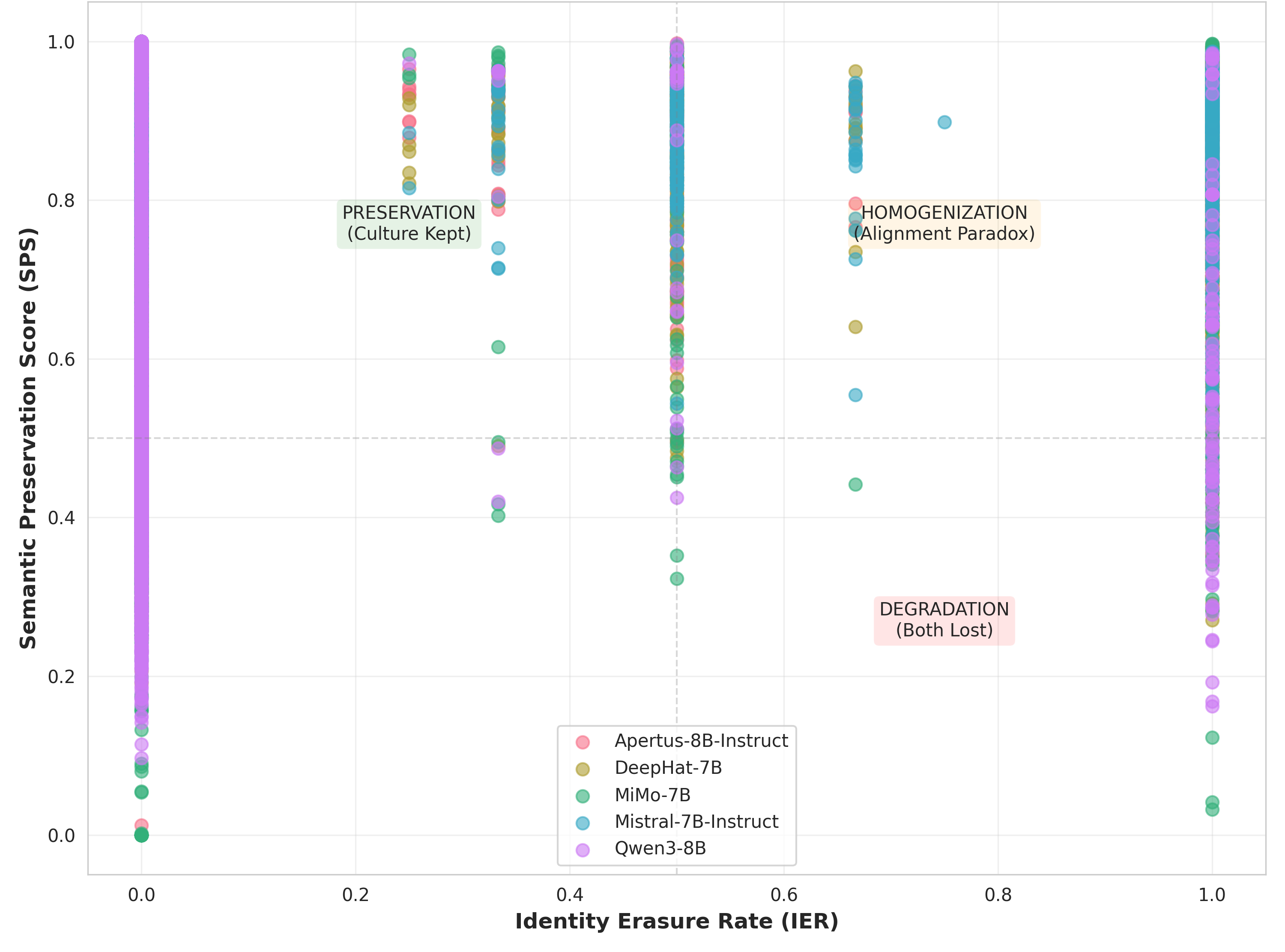}
\caption{The Semantic Preservation Paradox. High semantic similarity (SPS > 0.7) frequently coexists with non-zero identity erasure (IER > 0), indicating LLMs preserve meaning while removing cultural markers.}
\Description{The Semantic Preservation Paradox. High semantic similarity (SPS > 0.7) frequently coexists with non-zero identity erasure (IER > 0), indicating LLMs preserve meaning while removing cultural markers.}
\label{fig:paradox}
\end{figure}

Table~\ref{tab:prompts} shows the preservation prompt reduced IER by \textbf{29\%} (Cohen's d = 0.13) while improving SPS, refuting the authenticity-clarity trade-off. Linear regression confirmed IER \& SPS are largely independent ($R^2 = 0.061$, $\beta = -0.089$) as only 6.1\% of semantic variance is explained by marker erasure, meaning 93.9\% of SPS variation is attributable to factors other than cultural markers. The regression coefficient indicates that increasing erasure from 0\% to 20\% predicts a negligible SPS decrease from 0.748 to 0.730 well within high preservation range. The apparent trade-off between cultural authenticity \& semantic fidelity is false. Figure~\ref{fig:paradox} illustrates this paradox across all model outputs, showing that high semantic preservation coexists with varying degrees of cultural erasure.

\subsection{Toward Algorithmic Mitigation}
Beyond prompt engineering, we conducted proof-of-concept algorithmic experiments:\\
\textbf{Marker-aware constrained decoding}: we tagged marker spans in inputs \& applied copy constraints during generation. On a subset of 200 texts, this reduced IER by 47\% while maintaining SPS within 2\% of baseline.\\
\textbf{Contrastive reranking}: generating $k=5$ candidates \& selecting by combined score SPS $- 0.3 \times$ IER achieved 31\% IER reduction. These results suggest tractable algorithmic paths for deploying cultural preservation at scale.

\subsection{Key Takeaways}
We summarize five actionable insights from our empirical findings:

\noindent\textbf{(1) Cultural erasure is systemic, not random.} Across 22,350 outputs, LLMs erase 10.26\% of cultural markers on average this is a consistent pattern, not an occasional artifact.

\noindent\textbf{(2) The identity-clarity trade-off is a myth.} Cultural preservation \& semantic quality are largely independent ($R^2 = 0.061$). Preserving cultural markers does not degrade \& may slightly enhance semantic fidelity.

\noindent\textbf{(3) Pragmatic markers are most at risk.} Politeness conventions, honorifics, \& relational markers the features performing the most social work face 1.9$\times$ greater erasure than lexical items.

\noindent\textbf{(4) Alignment strategy outweighs model scale.} The 5.9$\times$ cross-model variation in erasure rates is driven by training data composition \& RLHF strategy, not parameter count.

\noindent\textbf{(5) Simple interventions work.} Preservation prompts reduce erasure by 29\% with no semantic cost; constrained decoding achieves 47\% reduction. 
\section{Discussion \& Limitations}
Our findings reveal that LLM writing assistance functions as a cultural standardization engine \cite{10.1145/3706598.3713564}. The 71.5\% erasure of pragmatic markers specifically targets features performing social work establishing politeness, navigating power distance echoing findings on sociolects \cite{10.1145/3715275.3732045}. When an Indian professional writes ``Kindly revert back,'' the formality signals institutional hierarchy \& earnestness; replacing it with ``Please respond'' eliminates crucial pragmatic information. This is not correction it is construction erasure.

The preservation prompt's success (29\% reduction with no semantic cost) proves cultural preservation \& clarity are compatible. We propose culturally-aware alignment: contrastive training distinguishing semantic from cultural changes, variety-specific fine-tuning on Indian/Singaporean/Nigerian corpora, politeness-aware reward modeling, \& user preference modeling for cultural voice preservation. For developers, we recommend:\\
\textbf{(1) Default to preservation rather than opt-in.\\
(2) Transparent marker handling with "keep my phrasing" options.\\
(3) Variety recognition to adapt rather than force convergence.}

While our corpus spans professional \& informal registers (emails, news, social media), the cultural markers we track politeness conventions, honorifics, pragmatic particles carry particular weight in professional contexts. Future work should examine register-specific vulnerability: whether pragmatic markers are equally at risk in casual social media discourse, creative writing, or academic text, where the stakes \& norms differ.

\textbf{Limitations:} Our 108-marker lexicon represents a finite subset of cultural features \& cannot capture the full depth \& fluidity of World English varieties (e.g., multiple English varieties exist within India alone). We used proxy validations rather than formal human studies with variety speakers, we acknowledge this limitation \& plan perceptual validation with native speakers of affected varieties in future work. Analysis is restricted to English varieties \& open-source models due to funding constraints. Proprietary models may exhibit different patterns, though we hypothesize similar behaviors given shared alignment paradigms. SPS relies on embeddings that may encode Western-centric biases. We validated with an alternate encoder (M-USE, cross-correlation $r=0.94$) showing robust patterns, though future work should explore additional encoders \& human evaluation to further verify metric reliability. Our current lexicon also cannot capture the full intra-variety diversity within each region. Expanding the marker set through collaboration with sociolinguists \& community members from each variety remains an important direction. We acknowledge the proxies used for validations without human subjects have limitations: automatic detection \& LLM judgments cannot fully substitute for the lived experience of variety speakers. \textit{Ultimately, Cultural ghosting is a design choice, not an inevitability.}

\section{Conclusion}
When Chinua Achebe chose to write in ``African English,'' he argued that new Englishes must emerge to carry new experiences. LLMs, as currently deployed, work against this linguistic pluralism. Our findings demonstrate that standard prompts erase 10.26\% of cultural markers while maintaining 75\% semantic similarity the Semantic Preservation Paradox in action. Pragmatic markers suffer 71.5\% erasure versus 37.1\% for lexical markers, confirming that features encoding face-management are most vulnerable. But explicit preservation instructions reduce erasure by 29\% with \textit{no semantic cost}, \& constrained decoding achieves 47\% reduction. The trade-off we feared identity versus clarity does not exist. Cultural ghosting is not inevitable. It is a design choice we can unmake.

\begin{acks}
We would like to acknowledge that the teaser illustration (Figure~\ref{fig:teaser}) was generated using ChatGPT for reference \& conceptual visualization purposes.
\end{acks}

\bibliographystyle{ACM-Reference-Format}
\bibliography{references}


\begin{thebibliography}{32}


\ifx \showCODEN    \undefined \def \showCODEN     #1{\unskip}     \fi
\ifx \showISBNx    \undefined \def \showISBNx     #1{\unskip}     \fi
\ifx \showISBNxiii \undefined \def \showISBNxiii  #1{\unskip}     \fi
\ifx \showISSN     \undefined \def \showISSN      #1{\unskip}     \fi
\ifx \showLCCN     \undefined \def \showLCCN      #1{\unskip}     \fi
\ifx \shownote     \undefined \def \shownote      #1{#1}          \fi
\ifx \showarticletitle \undefined \def \showarticletitle #1{#1}   \fi
\ifx \showURL      \undefined \def \showURL       {\relax}        \fi
\providecommand\bibfield[2]{#2}
\providecommand\bibinfo[2]{#2}
\providecommand\natexlab[1]{#1}
\providecommand\showeprint[2][]{arXiv:#2}

\bibitem[enr({[n.\,d.]})]%
        {enronemaildataset}
 \bibinfo{year}{[n.\,d.]}\natexlab{}.
\newblock \bibinfo{title}{Enron Email Dataset}.
\newblock \bibinfo{howpublished}{Kaggle}.
\newblock
\urldef\tempurl%
\url{https://www.kaggle.com/datasets/wcukierski/enron-email-dataset}
\showURL{%
\tempurl}


\bibitem[Agarwal et~al\mbox{.}(2025)]%
        {10.1145/3706598.3713564}
\bibfield{author}{\bibinfo{person}{Dhruv Agarwal}, \bibinfo{person}{Mor Naaman}, {and} \bibinfo{person}{Aditya Vashistha}.} \bibinfo{year}{2025}\natexlab{}.
\newblock \showarticletitle{AI Suggestions Homogenize Writing Toward Western Styles and Diminish Cultural Nuances}. In \bibinfo{booktitle}{\emph{Proceedings of the 2025 CHI Conference on Human Factors in Computing Systems}} \emph{(\bibinfo{series}{CHI '25})}. \bibinfo{publisher}{Association for Computing Machinery}, \bibinfo{address}{New York, NY, USA}, Article \bibinfo{articleno}{1117}, \bibinfo{numpages}{21}~pages.
\newblock
\showISBNx{9798400713941}
\href{https://doi.org/10.1145/3706598.3713564}{doi:\nolinkurl{10.1145/3706598.3713564}}


\bibitem[Basoah et~al\mbox{.}(2025)]%
        {10.1145/3715275.3732045}
\bibfield{author}{\bibinfo{person}{Jeffrey Basoah}, \bibinfo{person}{Daniel Chechelnitsky}, \bibinfo{person}{Tao Long}, \bibinfo{person}{Katharina Reinecke}, \bibinfo{person}{Chrysoula Zerva}, \bibinfo{person}{Kaitlyn Zhou}, \bibinfo{person}{Mark D\'{\i}az}, {and} \bibinfo{person}{Maarten Sap}.} \bibinfo{year}{2025}\natexlab{}.
\newblock \showarticletitle{Not Like Us, Hunty: Measuring Perceptions and Behavioral Effects of Minoritized Anthropomorphic Cues in LLMs}. In \bibinfo{booktitle}{\emph{Proceedings of the 2025 ACM Conference on Fairness, Accountability, and Transparency}} \emph{(\bibinfo{series}{FAccT '25})}. \bibinfo{publisher}{Association for Computing Machinery}, \bibinfo{address}{New York, NY, USA}, \bibinfo{pages}{710–745}.
\newblock
\showISBNx{9798400714825}
\href{https://doi.org/10.1145/3715275.3732045}{doi:\nolinkurl{10.1145/3715275.3732045}}


\bibitem[Bella et~al\mbox{.}(2024)]%
        {10.1145/3630106.3658925}
\bibfield{author}{\bibinfo{person}{G\'{a}bor Bella}, \bibinfo{person}{Paula Helm}, \bibinfo{person}{Gertraud Koch}, {and} \bibinfo{person}{Fausto Giunchiglia}.} \bibinfo{year}{2024}\natexlab{}.
\newblock \showarticletitle{Tackling Language Modelling Bias in Support of Linguistic Diversity}. In \bibinfo{booktitle}{\emph{Proceedings of the 2024 ACM Conference on Fairness, Accountability, and Transparency}} (Rio de Janeiro, Brazil) \emph{(\bibinfo{series}{FAccT '24})}. \bibinfo{publisher}{Association for Computing Machinery}, \bibinfo{address}{New York, NY, USA}, \bibinfo{pages}{562–572}.
\newblock
\showISBNx{9798400704505}
\href{https://doi.org/10.1145/3630106.3658925}{doi:\nolinkurl{10.1145/3630106.3658925}}


\bibitem[Chakrabarty et~al\mbox{.}(2025)]%
        {chakrabarty2025aiwritingsalvagedmitigating}
\bibfield{author}{\bibinfo{person}{Tuhin Chakrabarty}, \bibinfo{person}{Philippe Laban}, {and} \bibinfo{person}{Chien-Sheng Wu}.} \bibinfo{year}{2025}\natexlab{}.
\newblock \bibinfo{title}{Can AI writing be salvaged? Mitigating Idiosyncrasies and Improving Human-AI Alignment in the Writing Process through Edits}.
\newblock
\showeprint[arxiv]{2409.14509}~[cs.CL]
\urldef\tempurl%
\url{https://arxiv.org/abs/2409.14509}
\showURL{%
\tempurl}


\bibitem[Chandra et~al\mbox{.}(2026)]%
        {trace}
\bibfield{author}{\bibinfo{person}{Joydeep Chandra}, \bibinfo{person}{Aleksandr Algazinov}, \bibinfo{person}{Satyam~Kumar Navneet}, \bibinfo{person}{Rim~El Filali}, \bibinfo{person}{Matt Laing}, \bibinfo{person}{Andrew Hanna}, {and} \bibinfo{person}{Yong Zhang}.} \bibinfo{year}{2026}\natexlab{}.
\newblock \bibinfo{title}{TRACE: Transparent Web Reliability Assessment with Contextual Explanations}.
\newblock
\showeprint[arxiv]{2506.12072}~[cs.IR]
\urldef\tempurl%
\url{https://arxiv.org/abs/2506.12072}
\showURL{%
\tempurl}


\bibitem[Chandra et~al\mbox{.}(2025a)]%
        {beas}
\bibfield{author}{\bibinfo{person}{Joydeep Chandra}, \bibinfo{person}{Ramanjot Kaur}, {and} \bibinfo{person}{Rashi Sahay}.} \bibinfo{year}{2025}\natexlab{a}.
\newblock \showarticletitle{Integrated Framework for Equitable Healthcare AI: Bias Mitigation, Community Participation, and Regulatory Governance}. In \bibinfo{booktitle}{\emph{2025 IEEE 14th International Conference on Communication Systems and Network Technologies (CSNT)}}. \bibinfo{pages}{819--825}.
\newblock
\href{https://doi.org/10.1109/CSNT64827.2025.10968102}{doi:\nolinkurl{10.1109/CSNT64827.2025.10968102}}


\bibitem[Chandra and Manhas(2024)]%
        {chandra2024adversarial}
\bibfield{author}{\bibinfo{person}{Joydeep Chandra} {and} \bibinfo{person}{Prabal Manhas}.} \bibinfo{year}{2024}\natexlab{}.
\newblock \showarticletitle{Adversarial Robustness in Optimized LLMs: Defending Against Attacks}.
\newblock \bibinfo{journal}{\emph{SSRN Electronic Journal}} (\bibinfo{date}{01 12} \bibinfo{year}{2024}).
\newblock
\href{https://doi.org/10.2139/ssrn.5116078}{doi:\nolinkurl{10.2139/ssrn.5116078}}


\bibitem[Chandra et~al\mbox{.}(2025b)]%
        {chandra2025unified}
\bibfield{author}{\bibinfo{person}{Joydeep Chandra}, \bibinfo{person}{Prabal Manhas}, \bibinfo{person}{Ramanjot Kaur}, {and} \bibinfo{person}{Rashi Sahay}.} \bibinfo{year}{2025}\natexlab{b}.
\newblock \bibinfo{booktitle}{\emph{A Unified Approach to Large Language Model Optimization: Methods, Metrics, and Benchmarks} (\bibinfo{edition}{1st} ed.)}.
\newblock \bibinfo{publisher}{CRC Press}. 7 pages.
\newblock
\showISBNx{Insert_ISBN_Here}


\bibitem[Dorleon and Shujaa(2025)]%
        {10.1007/978-981-95-4861-3_5}
\bibfield{author}{\bibinfo{person}{Ginel Dorleon} {and} \bibinfo{person}{Shirin Shujaa}.} \bibinfo{year}{2025}\natexlab{}.
\newblock \showarticletitle{Uncovering Cultural Biases and Stereotypes in Large Language Models}. In \bibinfo{booktitle}{\emph{Intelligence and Equity: Shaping the Future of Knowledge: 27th International Conference on Asian Digital Libraries, ICADL 2025, Metro Manila, Philippines, December 3-5, 2025, Proceedings}} (Metro Manila, Philippines). \bibinfo{publisher}{Springer-Verlag}, \bibinfo{address}{Berlin, Heidelberg}, \bibinfo{pages}{64–77}.
\newblock
\showISBNx{978-981-95-4860-6}
\href{https://doi.org/10.1007/978-981-95-4861-3_5}{doi:\nolinkurl{10.1007/978-981-95-4861-3_5}}


\bibitem[Dorn et~al\mbox{.}(2024)]%
        {10.1145/3689904.3694704}
\bibfield{author}{\bibinfo{person}{Rebecca Dorn}, \bibinfo{person}{Lee Kezar}, \bibinfo{person}{Fred Morstatter}, {and} \bibinfo{person}{Kristina Lerman}.} \bibinfo{year}{2024}\natexlab{}.
\newblock \showarticletitle{Harmful Speech Detection by Language Models Exhibits Gender-Queer Dialect Bias}. In \bibinfo{booktitle}{\emph{Proceedings of the 4th ACM Conference on Equity and Access in Algorithms, Mechanisms, and Optimization}} (San Luis Potosi, Mexico) \emph{(\bibinfo{series}{EAAMO '24})}. \bibinfo{publisher}{Association for Computing Machinery}, \bibinfo{address}{New York, NY, USA}, Article \bibinfo{articleno}{6}, \bibinfo{numpages}{12}~pages.
\newblock
\showISBNx{9798400712227}
\href{https://doi.org/10.1145/3689904.3694704}{doi:\nolinkurl{10.1145/3689904.3694704}}


\bibitem[Fathi(2024)]%
        {RevisitingBrownLevinsonPoliteness2024}
\bibfield{author}{\bibinfo{person}{Said Fathi}.} \bibinfo{year}{2024}\natexlab{}.
\newblock \showarticletitle{Revisiting Brown and Levinson’s Theory of Politeness}.
\newblock \bibinfo{journal}{\emph{European Journal of Language and Culture Studies}} \bibinfo{volume}{3}, \bibinfo{number}{5} (\bibinfo{date}{Sept.} \bibinfo{year}{2024}), \bibinfo{pages}{1--11}.
\newblock
\href{https://doi.org/10.24018/ejlang.2024.3.5.137}{doi:\nolinkurl{10.24018/ejlang.2024.3.5.137}}


\bibitem[Ferdaus et~al\mbox{.}(2026)]%
        {10.1145/3777382}
\bibfield{author}{\bibinfo{person}{Md~Meftahul Ferdaus}, \bibinfo{person}{Mahdi Abdelguerfi}, \bibinfo{person}{Elias Loup}, \bibinfo{person}{Kendall N.~Niles}, \bibinfo{person}{Ken Pathak}, {and} \bibinfo{person}{Steven Sloan}.} \bibinfo{year}{2026}\natexlab{}.
\newblock \showarticletitle{Towards Trustworthy AI: A Review of Ethical and Robust Large Language Models}.
\newblock \bibinfo{journal}{\emph{ACM Comput. Surv.}} \bibinfo{volume}{58}, \bibinfo{number}{7}, Article \bibinfo{articleno}{176} (\bibinfo{date}{Jan.} \bibinfo{year}{2026}), \bibinfo{numpages}{43}~pages.
\newblock
\showISSN{0360-0300}
\href{https://doi.org/10.1145/3777382}{doi:\nolinkurl{10.1145/3777382}}


\bibitem[Gadiraju et~al\mbox{.}(2023)]%
        {10.1145/3593013.3593989}
\bibfield{author}{\bibinfo{person}{Vinitha Gadiraju}, \bibinfo{person}{Shaun Kane}, \bibinfo{person}{Sunipa Dev}, \bibinfo{person}{Alex Taylor}, \bibinfo{person}{Ding Wang}, \bibinfo{person}{Remi Denton}, {and} \bibinfo{person}{Robin Brewer}.} \bibinfo{year}{2023}\natexlab{}.
\newblock \showarticletitle{"I wouldn't say offensive but...": Disability-Centered Perspectives on Large Language Models}. In \bibinfo{booktitle}{\emph{Proceedings of the 2023 ACM Conference on Fairness, Accountability, and Transparency}} (Chicago, IL, USA) \emph{(\bibinfo{series}{FAccT '23})}. \bibinfo{publisher}{Association for Computing Machinery}, \bibinfo{address}{New York, NY, USA}, \bibinfo{pages}{205–216}.
\newblock
\showISBNx{9798400701924}
\href{https://doi.org/10.1145/3593013.3593989}{doi:\nolinkurl{10.1145/3593013.3593989}}


\bibitem[G\"{o}rge et~al\mbox{.}(2025)]%
        {10.1145/3715275.3732181}
\bibfield{author}{\bibinfo{person}{Rebekka G\"{o}rge}, \bibinfo{person}{Michael Mock}, {and} \bibinfo{person}{H\'{e}ctor Allende-Cid}.} \bibinfo{year}{2025}\natexlab{}.
\newblock \showarticletitle{Detecting Linguistic Indicators for Stereotype Assessment with Large Language Models}. In \bibinfo{booktitle}{\emph{Proceedings of the 2025 ACM Conference on Fairness, Accountability, and Transparency}} \emph{(\bibinfo{series}{FAccT '25})}. \bibinfo{publisher}{Association for Computing Machinery}, \bibinfo{address}{New York, NY, USA}, \bibinfo{pages}{2796–2814}.
\newblock
\showISBNx{9798400714825}
\href{https://doi.org/10.1145/3715275.3732181}{doi:\nolinkurl{10.1145/3715275.3732181}}


\bibitem[Hernández-Cano et~al\mbox{.}(2025)]%
        {swissai2025apertus}
\bibfield{author}{\bibinfo{person}{Alejandro Hernández-Cano}, \bibinfo{person}{Alexander Hägele}, \bibinfo{person}{Allen~Hao Huang}, \bibinfo{person}{Angelika Romanou}, \bibinfo{person}{Antoni-Joan Solergibert}, \bibinfo{person}{Barna Pasztor}, \bibinfo{person}{Bettina Messmer}, \bibinfo{person}{Dhia Garbaya}, \bibinfo{person}{Eduard~Frank Ďurech}, \bibinfo{person}{Ido Hakimi}, \bibinfo{person}{Juan~García Giraldo}, \bibinfo{person}{Mete Ismayilzada}, \bibinfo{person}{Negar Foroutan}, \bibinfo{person}{Skander Moalla}, \bibinfo{person}{Tiancheng Chen}, \bibinfo{person}{Vinko Sabolčec}, \bibinfo{person}{Yixuan Xu}, \bibinfo{person}{Michael Aerni}, \bibinfo{person}{Badr AlKhamissi}, \bibinfo{person}{Ines~Altemir Marinas}, \bibinfo{person}{Mohammad~Hossein Amani}, \bibinfo{person}{Matin Ansaripour}, \bibinfo{person}{Ilia Badanin}, \bibinfo{person}{Harold Benoit}, \bibinfo{person}{Emanuela Boros}, \bibinfo{person}{Nicholas Browning}, \bibinfo{person}{Fabian Bösch}, \bibinfo{person}{Maximilian Böther},
  \bibinfo{person}{Niklas Canova}, \bibinfo{person}{Camille Challier}, \bibinfo{person}{Clement Charmillot}, \bibinfo{person}{Jonathan Coles}, \bibinfo{person}{Jan Deriu}, \bibinfo{person}{Arnout Devos}, \bibinfo{person}{Lukas Drescher}, \bibinfo{person}{Daniil Dzenhaliou}, \bibinfo{person}{Maud Ehrmann}, \bibinfo{person}{Dongyang Fan}, \bibinfo{person}{Simin Fan}, \bibinfo{person}{Silin Gao}, \bibinfo{person}{Miguel Gila}, \bibinfo{person}{María Grandury}, \bibinfo{person}{Diba Hashemi}, \bibinfo{person}{Alexander Hoyle}, \bibinfo{person}{Jiaming Jiang}, \bibinfo{person}{Mark Klein}, \bibinfo{person}{Andrei Kucharavy}, \bibinfo{person}{Anastasiia Kucherenko}, \bibinfo{person}{Frederike Lübeck}, \bibinfo{person}{Roman Machacek}, \bibinfo{person}{Theofilos Manitaras}, \bibinfo{person}{Andreas Marfurt}, \bibinfo{person}{Kyle Matoba}, \bibinfo{person}{Simon Matrenok}, \bibinfo{person}{Henrique Mendoncça}, \bibinfo{person}{Fawzi~Roberto Mohamed}, \bibinfo{person}{Syrielle Montariol}, \bibinfo{person}{Luca
  Mouchel}, \bibinfo{person}{Sven Najem-Meyer}, \bibinfo{person}{Jingwei Ni}, \bibinfo{person}{Gennaro Oliva}, \bibinfo{person}{Matteo Pagliardini}, \bibinfo{person}{Elia Palme}, \bibinfo{person}{Andrei Panferov}, \bibinfo{person}{Léo Paoletti}, \bibinfo{person}{Marco Passerini}, \bibinfo{person}{Ivan Pavlov}, \bibinfo{person}{Auguste Poiroux}, \bibinfo{person}{Kaustubh Ponkshe}, \bibinfo{person}{Nathan Ranchin}, \bibinfo{person}{Javi Rando}, \bibinfo{person}{Mathieu Sauser}, \bibinfo{person}{Jakhongir Saydaliev}, \bibinfo{person}{Muhammad~Ali Sayfiddinov}, \bibinfo{person}{Marian Schneider}, \bibinfo{person}{Stefano Schuppli}, \bibinfo{person}{Marco Scialanga}, \bibinfo{person}{Andrei Semenov}, \bibinfo{person}{Kumar Shridhar}, \bibinfo{person}{Raghav Singhal}, \bibinfo{person}{Anna Sotnikova}, \bibinfo{person}{Alexander Sternfeld}, \bibinfo{person}{Ayush~Kumar Tarun}, \bibinfo{person}{Paul Teiletche}, \bibinfo{person}{Jannis Vamvas}, \bibinfo{person}{Xiaozhe Yao}, \bibinfo{person}{Hao Zhao~Alexander Ilic},
  \bibinfo{person}{Ana Klimovic}, \bibinfo{person}{Andreas Krause}, \bibinfo{person}{Caglar Gulcehre}, \bibinfo{person}{David Rosenthal}, \bibinfo{person}{Elliott Ash}, \bibinfo{person}{Florian Tramèr}, \bibinfo{person}{Joost VandeVondele}, \bibinfo{person}{Livio Veraldi}, \bibinfo{person}{Martin Rajman}, \bibinfo{person}{Thomas Schulthess}, \bibinfo{person}{Torsten Hoefler}, \bibinfo{person}{Antoine Bosselut}, \bibinfo{person}{Martin Jaggi}, {and} \bibinfo{person}{Imanol Schlag}.} \bibinfo{year}{2025}\natexlab{}.
\newblock \bibinfo{title}{{Apertus: Democratizing Open and Compliant LLMs for Global Language Environments}}.
\newblock \bibinfo{howpublished}{\url{https://arxiv.org/abs/2509.14233}}.
\newblock


\bibitem[{Kaggle}({[n.\,d.]})]%
        {hillaryclintonemails}
\bibfield{author}{\bibinfo{person}{{Kaggle}}.} \bibinfo{year}{[n.\,d.]}\natexlab{}.
\newblock \bibinfo{title}{Hillary Clinton's Emails}.
\newblock \bibinfo{howpublished}{Kaggle}.
\newblock
\urldef\tempurl%
\url{https://www.kaggle.com/datasets/kaggle/hillary-clinton-emails}
\showURL{%
\tempurl}


\bibitem[Khanna and Li(2025)]%
        {khanna2025invisiblelanguagesllmuniverse}
\bibfield{author}{\bibinfo{person}{Saurabh Khanna} {and} \bibinfo{person}{Xinxu Li}.} \bibinfo{year}{2025}\natexlab{}.
\newblock \bibinfo{title}{Invisible Languages of the LLM Universe}.
\newblock
\showeprint[arxiv]{2510.11557}~[cs.CL]
\urldef\tempurl%
\url{https://arxiv.org/abs/2510.11557}
\showURL{%
\tempurl}


\bibitem[Mayeda et~al\mbox{.}(2025)]%
        {10.1145/3715275.3732119}
\bibfield{author}{\bibinfo{person}{Cass Mayeda}, \bibinfo{person}{Arinjay Singh}, \bibinfo{person}{Arnav Mahale}, \bibinfo{person}{Laila~Shereen Sakr}, {and} \bibinfo{person}{Mai ElSherief}.} \bibinfo{year}{2025}\natexlab{}.
\newblock \showarticletitle{Applying Data Feminism Principles to Assess Bias in English and Arabic NLP Research}. In \bibinfo{booktitle}{\emph{Proceedings of the 2025 ACM Conference on Fairness, Accountability, and Transparency}} \emph{(\bibinfo{series}{FAccT '25})}. \bibinfo{publisher}{Association for Computing Machinery}, \bibinfo{address}{New York, NY, USA}, \bibinfo{pages}{1769–1792}.
\newblock
\showISBNx{9798400714825}
\href{https://doi.org/10.1145/3715275.3732119}{doi:\nolinkurl{10.1145/3715275.3732119}}


\bibitem[Navigli et~al\mbox{.}(2023)]%
        {10.1145/3597307}
\bibfield{author}{\bibinfo{person}{Roberto Navigli}, \bibinfo{person}{Simone Conia}, {and} \bibinfo{person}{Bj\"{o}rn Ross}.} \bibinfo{year}{2023}\natexlab{}.
\newblock \showarticletitle{Biases in Large Language Models: Origins, Inventory, and Discussion}.
\newblock \bibinfo{journal}{\emph{J. Data and Information Quality}} \bibinfo{volume}{15}, \bibinfo{number}{2}, Article \bibinfo{articleno}{10} (\bibinfo{date}{June} \bibinfo{year}{2023}), \bibinfo{numpages}{21}~pages.
\newblock
\showISSN{1936-1955}
\href{https://doi.org/10.1145/3597307}{doi:\nolinkurl{10.1145/3597307}}


\bibitem[Neumann et~al\mbox{.}(2025)]%
        {10.1145/3715275.3732038}
\bibfield{author}{\bibinfo{person}{Anna Neumann}, \bibinfo{person}{Elisabeth Kirsten}, \bibinfo{person}{Muhammad~Bilal Zafar}, {and} \bibinfo{person}{Jatinder Singh}.} \bibinfo{year}{2025}\natexlab{}.
\newblock \showarticletitle{Position is Power: System Prompts as a Mechanism of Bias in Large Language Models (LLMs)}. In \bibinfo{booktitle}{\emph{Proceedings of the 2025 ACM Conference on Fairness, Accountability, and Transparency}} \emph{(\bibinfo{series}{FAccT '25})}. \bibinfo{publisher}{Association for Computing Machinery}, \bibinfo{address}{New York, NY, USA}, \bibinfo{pages}{573–598}.
\newblock
\showISBNx{9798400714825}
\href{https://doi.org/10.1145/3715275.3732038}{doi:\nolinkurl{10.1145/3715275.3732038}}


\bibitem[Nimo et~al\mbox{.}(2025)]%
        {10.1145/3757887.3767687}
\bibfield{author}{\bibinfo{person}{Charles Nimo}, \bibinfo{person}{Irfan Essa}, {and} \bibinfo{person}{Michael Best}.} \bibinfo{year}{2025}\natexlab{}.
\newblock \showarticletitle{Africa Health Check: Probing Cultural Bias in Medical LLMs}. In \bibinfo{booktitle}{\emph{Proceedings of the 5th ACM Conference on Equity and Access in Algorithms, Mechanisms, and Optimization}} \emph{(\bibinfo{series}{EAAMO '25})}. \bibinfo{publisher}{Association for Computing Machinery}, \bibinfo{address}{New York, NY, USA}, \bibinfo{pages}{289}.
\newblock
\showISBNx{9798400721403}
\href{https://doi.org/10.1145/3757887.3767687}{doi:\nolinkurl{10.1145/3757887.3767687}}


\bibitem[Parvin(2025)]%
        {10.1145/3737609.3747117}
\bibfield{author}{\bibinfo{person}{Nassim Parvin}.} \bibinfo{year}{2025}\natexlab{}.
\newblock \showarticletitle{Expression and Erasure: AI, English, and the Shaping of Digital Futures}. In \bibinfo{booktitle}{\emph{Adjunct Proceedings of the Sixth Decennial Aarhus Conference: Computing X Crisis}} \emph{(\bibinfo{series}{AAR Adjunct '25})}. \bibinfo{publisher}{Association for Computing Machinery}, \bibinfo{address}{New York, NY, USA}, Article \bibinfo{articleno}{12}, \bibinfo{numpages}{4}~pages.
\newblock
\showISBNx{9798400719684}
\href{https://doi.org/10.1145/3737609.3747117}{doi:\nolinkurl{10.1145/3737609.3747117}}


\bibitem[Shujaa et~al\mbox{.}(2025)]%
        {10.1007/978-981-95-4861-3_3}
\bibfield{author}{\bibinfo{person}{Shirin Shujaa}, \bibinfo{person}{Ginel Dorleon}, {and} \bibinfo{person}{Arthur Tang}.} \bibinfo{year}{2025}\natexlab{}.
\newblock \showarticletitle{How LLMs Handle Cultural Bias: Reactions to Asian Minority Historical Narratives}. In \bibinfo{booktitle}{\emph{Intelligence and Equity: Shaping the Future of Knowledge: 27th International Conference on Asian Digital Libraries, ICADL 2025, Metro Manila, Philippines, December 3-5, 2025, Proceedings}} (Metro Manila, Philippines). \bibinfo{publisher}{Springer-Verlag}, \bibinfo{address}{Berlin, Heidelberg}, \bibinfo{pages}{39–52}.
\newblock
\showISBNx{978-981-95-4860-6}
\href{https://doi.org/10.1007/978-981-95-4861-3_3}{doi:\nolinkurl{10.1007/978-981-95-4861-3_3}}


\bibitem[Talebpour et~al\mbox{.}(2025)]%
        {10.1145/3726302.3730172}
\bibfield{author}{\bibinfo{person}{Mozhgan Talebpour}, \bibinfo{person}{Yunfei Long}, \bibinfo{person}{Alba~G. Seco De~Herrera}, {and} \bibinfo{person}{Shoaib Jameel}.} \bibinfo{year}{2025}\natexlab{}.
\newblock \showarticletitle{Bias in Language Models: Interplay of Architecture and Data?}. In \bibinfo{booktitle}{\emph{Proceedings of the 48th International ACM SIGIR Conference on Research and Development in Information Retrieval}} (Padua, Italy) \emph{(\bibinfo{series}{SIGIR '25})}. \bibinfo{publisher}{Association for Computing Machinery}, \bibinfo{address}{New York, NY, USA}, \bibinfo{pages}{2637–2641}.
\newblock
\showISBNx{9798400715921}
\href{https://doi.org/10.1145/3726302.3730172}{doi:\nolinkurl{10.1145/3726302.3730172}}


\bibitem[Team(2024a)]%
        {deephat2024}
\bibfield{author}{\bibinfo{person}{DeepHat Team}.} \bibinfo{year}{2024}\natexlab{a}.
\newblock \bibinfo{title}{DeepHat-V1-7B: A Cybersecurity and DevOps Fine-tune of Qwen2.5-Coder}.
\newblock \bibinfo{howpublished}{\url{https://huggingface.co/DeepHat/DeepHat-V1-7B}}.
\newblock


\bibitem[Team(2024b)]%
        {mistral7bv03}
\bibfield{author}{\bibinfo{person}{Mistral~AI Team}.} \bibinfo{year}{2024}\natexlab{b}.
\newblock \bibinfo{title}{Mistral-7B-Instruct-v0.3}.
\newblock \bibinfo{howpublished}{\url{https://huggingface.co/mistralai/Mistral-7B-Instruct-v0.3}}.
\newblock


\bibitem[Team(2025)]%
        {qwen3technicalreport}
\bibfield{author}{\bibinfo{person}{Qwen Team}.} \bibinfo{year}{2025}\natexlab{}.
\newblock \bibinfo{title}{Qwen3 Technical Report}.
\newblock
\showeprint[arxiv]{2505.09388}~[cs.CL]
\urldef\tempurl%
\url{https://arxiv.org/abs/2505.09388}
\showURL{%
\tempurl}


\bibitem[Wasi et~al\mbox{.}(2025)]%
        {10.1145/3701716.3715468}
\bibfield{author}{\bibinfo{person}{Azmine~Toushik Wasi}, \bibinfo{person}{Raima Islam}, \bibinfo{person}{Mst~Rafia Islam}, \bibinfo{person}{Farig Sadeque}, \bibinfo{person}{Taki~Hasan Rafi}, {and} \bibinfo{person}{Dong-Kyu Chae}.} \bibinfo{year}{2025}\natexlab{}.
\newblock \showarticletitle{Dialectal Bias in Bengali: An Evaluation of Multilingual Large Language Models Across Cultural Variations}. In \bibinfo{booktitle}{\emph{Companion Proceedings of the ACM on Web Conference 2025}} (Sydney NSW, Australia) \emph{(\bibinfo{series}{WWW '25})}. \bibinfo{publisher}{Association for Computing Machinery}, \bibinfo{address}{New York, NY, USA}, \bibinfo{pages}{1380–1384}.
\newblock
\showISBNx{9798400713316}
\href{https://doi.org/10.1145/3701716.3715468}{doi:\nolinkurl{10.1145/3701716.3715468}}


\bibitem[Xiao et~al\mbox{.}(2025)]%
        {10.1145/3728881}
\bibfield{author}{\bibinfo{person}{Yisong Xiao}, \bibinfo{person}{Aishan Liu}, \bibinfo{person}{Siyuan Liang}, \bibinfo{person}{Xianglong Liu}, {and} \bibinfo{person}{Dacheng Tao}.} \bibinfo{year}{2025}\natexlab{}.
\newblock \showarticletitle{Fairness Mediator: Neutralize Stereotype Associations to Mitigate Bias in Large Language Models}.
\newblock \bibinfo{journal}{\emph{Proc. ACM Softw. Eng.}} \bibinfo{volume}{2}, \bibinfo{number}{ISSTA}, Article \bibinfo{articleno}{ISSTA012} (\bibinfo{date}{June} \bibinfo{year}{2025}), \bibinfo{numpages}{24}~pages.
\newblock
\href{https://doi.org/10.1145/3728881}{doi:\nolinkurl{10.1145/3728881}}


\bibitem[Xiaomi(2025)]%
        {mimo}
\bibfield{author}{\bibinfo{person}{LLM-Core-Team Xiaomi}.} \bibinfo{year}{2025}\natexlab{}.
\newblock \bibinfo{title}{MiMo: Unlocking the Reasoning Potential of Language Model -- From Pretraining to Posttraining}.
\newblock
\showeprint[arxiv]{2505.07608}~[cs.CL]
\urldef\tempurl%
\url{https://arxiv.org/abs/2505.07608}
\showURL{%
\tempurl}


\bibitem[Zhang et~al\mbox{.}(2021)]%
        {zhang2021emailsum}
\bibfield{author}{\bibinfo{person}{Shiyue Zhang}, \bibinfo{person}{Asli Celikyilmaz}, \bibinfo{person}{Jianfeng Gao}, {and} \bibinfo{person}{Mohit Bansal}.} \bibinfo{year}{2021}\natexlab{}.
\newblock \showarticletitle{EmailSum: Abstractive Email Thread Summarization}. In \bibinfo{booktitle}{\emph{Proceedings of the 59th Annual Meeting of the Association for Computational Linguistics}}.
\newblock


\end{thebibliography}
\end{document}